\documentclass{article}

\usepackage{arxiv}

\usepackage[utf8]{inputenc} 
\usepackage[T1]{fontenc}    
\usepackage{hyperref}       
\usepackage{url}            
\usepackage{booktabs}       
\usepackage{amsfonts}       
\usepackage{nicefrac}       
\usepackage{microtype}      
\usepackage{lipsum}
\usepackage{graphicx}
\usepackage{authblk}
\usepackage{float}
\graphicspath{ {./images/} }

\usepackage{amsmath,amssymb}
\usepackage{comment}
\usepackage{mathrsfs}
\usepackage{graphicx,nicefrac}
\usepackage{float}

\usepackage{comment}
\usepackage{scalerel}
\usepackage[overload]{empheq}
\usepackage{textgreek}

\providecommand{\vx}{\mathbf{x}}
\providecommand{\vy}{\mathbf{y}}

\providecommand{\R}{\ensuremath{\mathbb{R}}}

\usepackage{algorithm}
\usepackage{algpseudocode}
\usepackage{enumitem}
\usepackage{hyperref}
\usepackage{mathrsfs}
\usepackage{graphicx,nicefrac}

\usepackage[utf8]{inputenc}
\usepackage[english]{babel}

\usepackage[margin=1cm]{caption}
\usepackage{comment}
\usepackage{scalerel}
\usepackage[overload]{empheq}

\usepackage{hyperref}

\begin{document}

\pagestyle{fancy}
\rhead{}

\title{Imaging dynamics beneath turbid media via parallelized single-photon detection}

\author[1]{Shiqi Xu}
\author[1]{Xi Yang}
\author[1,2]{Wenhui Liu}
\author[3]{Joakim Jonsson}
\author[1]{Ruobing Qian}
\author[1]{Pavan Chandra Konda}
\author[1]{Kevin C. Zhou}
\author[1]{Lucas Kreiss}
\author[2]{Qionghai Dai}
\author[4]{Haoqian Wang}
\author[3]{Edouard Berrocal}
\author[1,5,6,*]{Roarke Horstmeyer}

\affil[1]{Department of Biomedical Engineering, Duke University, Durham, NC, USA, 27708}
\affil[2]{Department of Automation, Tsinghua University, Beijing, China, 100084}
\affil[3]{Division of Combustion Physics, Department of Physics, Lund University, Sweden,22100}
\affil[4]{Tsinghua Shenzhen International Graduate School, Tsinghua University, Shenzhen, China, 518055}
\affil[5]{Department of Electrical and Computer Engineering, Duke University, Durham, NC, USA, 27708}
\affil[6]{Department of Physics, Duke University, Durham, NC, USA, 27708}
\affil[*]{roarke.w.horstmeyer@duke.edu}

\maketitle







\begin{abstract}
Noninvasive optical imaging through dynamic scattering media has numerous important biomedical applications but still remains a challenging task. While standard diffuse imaging methods measure optical absorption or fluorescent emission, it is also well-established that the temporal correlation of scattered coherent light diffuses through tissue much like optical intensity. Few works to date, however, have aimed to experimentally measure and process such temporal correlation data to demonstrate deep-tissue video reconstruction of decorrelation dynamics. In this work, we utilize a single-photon avalanche diode (SPAD) array camera to simultaneously monitor the temporal dynamics of speckle fluctuations at the single-photon level from 12 different phantom tissue surface locations delivered via a customized fiber bundle array. We then apply a deep neural network to convert the acquired single-photon measurements into video of scattering dynamics beneath rapidly decorrelating tissue phantoms. We demonstrate the ability to reconstruct images of transient (0.1-0.4s) dynamic events occurring up to 8 mm beneath a decorrelating tissue phantom with millimeter-scale resolution, and highlight how our model can flexibly extend to monitor flow speed within buried phantom vessels.

\end{abstract}


\section{Introduction}
\label{sec:intro}
Imaging deep within human tissue is a central challenge in biomedical optics. Over the past several decades, a wide variety of approaches have been developed to address this challenge at various scales. These include confocal~\cite{ntziachristos2010going} and non-linear~\cite{horton2013vivo} microscopy techniques that can image up to one millimeter deep within tissue, as well as novel wavefront shaping~\cite{horstmeyer2015guidestar}, time-of-flight diffuse optics~\cite{lyons2019computational,lindell2020three}, and photoacoustic techniques~\cite{wang2012photoacoustic} that can extend imaging depths to centimeter scales at reduced resolution. While there are many experimental demonstrations of imaging through thick scattering material, only a few of these techniques can easily be translated to living tissue specifically, or to dynamic scattering media in general. Dynamic scattering specimens, such as tissue decorrelate~\cite{jang2015relation} - microscopic movements due to effects like thermal variations and cell migration, for example, cause the optical scattering signature of a particular specimen to change rapidly over time. This rapid movement often presents challenges to effective \emph{in-vivo} deep-tissue imaging. While prior wavefront shaping methods can overcome such effects to focus within thick tissue at high speeds ~\cite{wang2015focusing,tzang2019wavefront,ruan2020fluorescence}, significant engineering challenges remain to achieve deep-tissue imaging in human subjects~\cite{gigan2021roadmap}. 

Instead of attempting to avoid or overcome the effects of decorrelation on imaging measurements, one alternative strategy is to directly measure such dynamic changes within the scattering specimens, and use these changes to aid with image formation. Here, the primary goal is \emph{not} to form intensity-based images, as in absorption or fluorescence microscopy, but to create a spatial map of fluctuation. This is typically achieved by measuring the temporal dynamics (e.g., temporal variance or correlation) of scattered radiation. Several important biological phenomena cause such temporal variation of an optical field, ranging from blood flow to neuronal firing events~\cite{korolevich2000experimental,dunn2001dynamic,culver2003diffuse,kim2003intracellular}. Optical coherence tomography angiography~\cite{wang2007three}, laser speckle contrast imaging~\cite{dunn2001dynamic}, as well as photoacoustic Doppler microscope ~\cite{yao2010vivo} have been developed to image such dynamics close to the tissue surface. However, to detect an optical signal that has traveled deep inside living tissue, which increasingly attenuates and decorrelates the optical field, one typically needs to eventually rely on fast single-photon-sensitive detection techniques that record optical fluctuations at approximately mega-hertz rates.

\begin{figure}[ht!]
\begin{center}
\includegraphics[width=15cm]{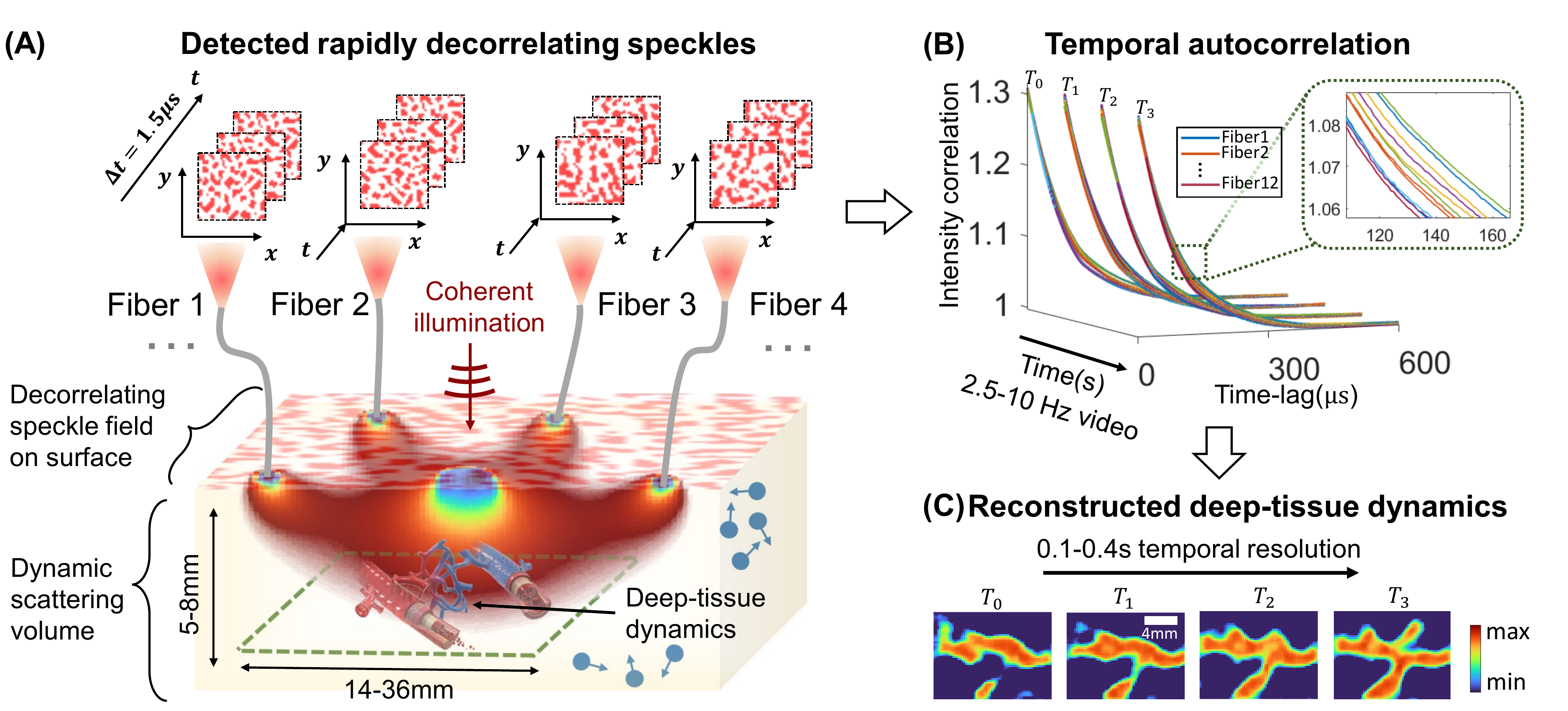}
\end{center}
\caption{Flow diagram of proposed method for imaging temporal decorrelation dynamics. (A) Illustration of Parallelized Diffuse Correlation Imaging (PaDI) measurement strategy. Scattered coherent light from source to multiple detector fibers travels through decorrelating scattering media along unique banana-shaped paths. Fully developed speckle on the tissue surface rapidly fluctuates as a function of deep-tissue movement. Green dashed box marks deep-tissue dynamics areas of interest for imaging. (B) Computed autocorrelation curves from time-resolved measurements of surface speckle at different tissue surface locations. (C) Autocorrelation variations caused by deep-tissue dynamics are computationally mapped into spatially resolved images of transient dynamics.}
\label{fig:fig1_flow}
\end{figure}

One established technique to detect dynamic scattering multiple centimeters within deep tissue is termed diffuse correlation spectroscopy(DCS)~\cite{durduran2014diffuse}, which records coherent light fluctuations. When coherent light enters a turbid medium, it randomly scatters and produces speckle. Movements within the tissue volume (e.g., cellular movement or blood flow) occur at different spatial locations and interact with the scattered optical field. By measuring temporal fluctuations of the scattered light at the tissue surface, it is possible to estimate a spatiotemporal map of decorrelating events. While such methods are widely used to assess blood flow variations across finite tissue areas as deep as beneath the adult skull ~\cite{buckley2014diffuse}, there has been limited work to date to rapidly form spatially resolved \emph{images} and \emph{video} of dynamic events beneath turbid media~\cite{durduran2014diffuse}, despite early work demonstrating that the temporal correlation of light transports through tissue follows a well-known diffusion process~\cite{boas1995scattering}. Three main challenges have prevented imaging of deep-tissue dynamics: 1) a low signal-to-noise (SNR) due to a limited number of available photons at requisite measurement rates, 2) a limited number of detectors to collect light from different locations across the scatterer surface, and 3) a challenging ill-posed inverse problem to map acquired data to accurate imagery.

To solve the first two challenges listed above, this work uses a single-photon avalanche diode (SPAD) array to simultaneously measure speckle field fluctuations across the tissue surface at the requisite sampling rates ($\sim$\textmu s) and single-photon sensitivities needed for deep detection~\cite{bruschini2019single}. Recently developed SPAD arrays, based on standard complementary metal-oxide-semiconductor(CMOS) fabrication technology, can integrate up to a million SPAD pixels onto a small chip~\cite{morimoto2020megapixel,canon2021mega}. This led to new imaging applications in fluorescence lifetime imaging~\cite{zickus2020wide}, scanning microscopy~\cite{buttafava2020spad}, confocal fluorescence fluctuation spectroscopy~\cite{slenders2021confocal}, Fourier ptychography~\cite{Xi2021quan}, as well as computer vision tasks, such as depth profile estimation~\cite{morimoto2020megapixel,gyongy2020high}, seeing around corners~\cite{gariepy2016detection} and through scattering slabs~\cite{satat2016all,lyons2019computational}. Most prior DCS measurement systems relied on fast single-pixel single-photon detectors (including single-pixel SPAD and photomultiplier tubes) for optical measurement ~\cite{durduran2014diffuse}. Single-pixel strategies for DCS-based image formation have several fundamental limitations. While several works demonstrated DCS-based imaging of temporal correlations in the past ~\cite{boas1995scattering,culver2003diffuse,zhou2006diffuse,han2015non,he2015noncontact}, none simultaneously acquired DCS signal from multiple tissue surface areas, as required for rapid image formation (e.g. to avoid effects of subject movement). Instead, these prior works mechanically scanned the specimen, or illumination and detection locations in a step-and-repeat fashion to measure speckle from different surface locations on a single detector. Furthermore, as only one or a few speckle modes can be sampled by a single detector while still maintaining suitable contrast, a long (seconds or more) measurement sequence is typically required to obtain a suitable signal-to-noise ratio for each measured temporal correlation curve (i.e., each surface location). This limited correlation measurement rate is quite detrimental - it precludes observation of dynamic variations of the subject pulse signal, for example, which can vary at sub-hertz rates. Recent work has demonstrated how parallelized speckle detection across many optical sensor pixels ~\cite{johansson2019multipixel,liu2020classifying,liu2021fast,sie2020high,xu2021rapid,zhou2021functional,xu2021diffusing,xu2022speckle,xu2022transient} can lead to significantly faster correlation sampling rates. We build upon these insights to create a new system capable of recording spatially resolved videos of temporal decorrelation without any moving parts. 

The third challenge noted above relates to the computational formation of dynamic images from limited measurement locations across the scatterer surface, typically formulated as an ill-posed inverse diffusion problem. While model-based solvers have demonstrated effective dynamics imaging in prior work~\cite{boas1995scattering,culver2003diffuse,zhou2006diffuse,han2015non,he2015noncontact}, simple scattering geometries were typically assumed (e.g., infinite and semi-infinite geometries). To alleviate model-based reconstruction issues, one can adopt a data-driven image reconstruction approach. Typically formed via training of a non-linear estimator with large amount of labeled data, neural network-based models have been used in the past to image static amplitude or phase objects through and within scattering medium using both all-optics ~\cite{horisaki2016learning,li2018deep,li2018imaging,rahmani2018multimode,lyu2019learning,sun2021scalable} and photoacoustic methods~\cite{gao2021deep}.  Inspired by such recent progress, we have developed a system and data post-processing pipeline, termed Parallelized Diffuse Correlation Imaging(PaDI), that addresses the above challenges to form images and video of transient dynamics events beneath multiple millimeters of decorrelating turbid media. Our new optical probe can image within a 140 mm$^2$ field-of-view at 5-8 mm depths beneath a decorrelating liquid tissue phantom ($\mu_a=0.01$ mm$^{-1}$, $\mu_s'=0.7$ mm$^{-1}$, Brownian coefficient $D=1.5\times10^6$ mm$^2$, for example - although many of these parameters can be flexibly adjusted) without any moving parts at multi-hertz video frame rate.

\section{Results and Discussion}

\subsection{Parallelized Diffuse Correlation Imaging (PaDI)}
\label{subsec::exp_setup}
\begin{figure}[!t]
\begin{center}
\includegraphics[width=15.0cm]{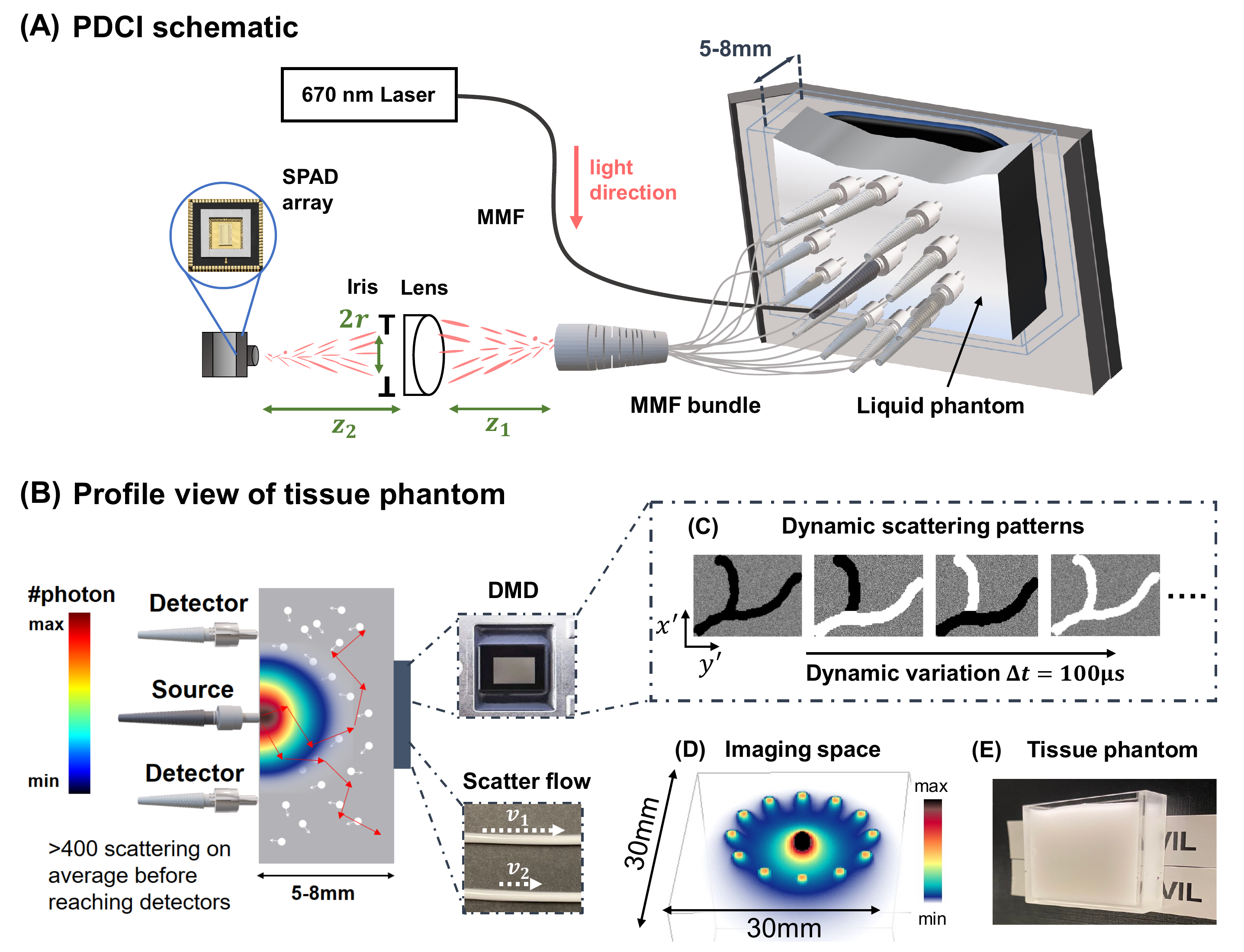}
\end{center}
\caption{(A) Schematic of PaDI system for imaging decorrelation. Back-scattered coherent light from single input port is collected by 12 multimode fibers (MMF) at tissue phantom surface and guided to SPAD array camera. (B) Profile view of the  tissue phantom imaging experiment. Digital micro-mirror device (DMD) and vessel phantom serve as source of temporal dynamics and is hidden beneath phantom by placing it immediately adjacent (separated by coverglass). All sources and detectors are placed on the same side of phantom. Colormap provides qualitative photon distribution map, where quantitative plot of sub-surface photon distribution is in Supplement Fig.1(B). (C) A set of DMD patterns that can be used to generate spatiotemporal varying dynamics. (D) Simulation of photon-sensitive region of our 12-fiber system. (E) A picture of the tissue phantom we use in experiments.}
\label{fig:fig2_sche}
\end{figure}

The phantom design and imaging setup is outlined in Fig. \ref{fig:fig2_sche}. To assess the performance of our PaDI system, we turn to an easily re-configurable non-biological liquid phantom setup that offers the ability to flexibly generate unique image targets with known spatial and temporal properties. To mimic decorrelation rates and scattering properties of human tissue, we utilized a liquid phantom filled with $1\mu\text{m}$-diameter polystyrene microspheres ($4.55\times10^6$\#/mm$^3$) solution enclosed in a custom-designed thin-walled cuvette as rapidly decorrelating turbid volume to occlude the target of interest. The target exhibits a reduced scattering coefficient of $0.7\text{mm}^{-1}$ as computed by the Lorenz-Mie method, and an experimentally measured absorption coefficient of $0.01\text{mm}^{-1}$. Also, based on fitting using a Monte Carlo method~\cite{boas2016establishing}, the medium exhibits an estimated Brownian motion diffuse constant of $1.5\times10^6\text{mm}^2$, which is close to the diffusion coefficient measured in model organisms ~\cite{durduran2010diffuse}. Supplement Section 3 details how these values are estimated. To generate expected temporal fluctuation variations within living tissue caused e.g. by blood flow, we placed a digital micro-mirror device (DMD) immediately behind this tissue phantom, with which we computationally created spatiotemporally varying patterns at kilohertz rates~\cite{liu2021fast}. Further, for the second generalizability study discussed in the \emph{experimental validation}, we also place two plastic tubes containing flowing scattering liquid with the same optical properties as the background volume. The movement of the liquid inside the tube is controlled with two syringe pumps (New Era, US1010).

Our light source is a 670nm diode-pumped solid-state (DPSS) laser (MSL-FN-671, Opto Engine LLC, USA) with a coherence length $\geq10$m, which we attenuated to 200 mW to match standard ANSI safety limits for illuminating tissue with visible light ~\cite{american2007american}. We guided this light to the liquid phantom surface using a $50\mu\text{m}$, $0.22$ numerical aperture (NA) multi-mode fiber (MMF). Before the MMF, we ensured that the DPSS laser output was effectively a single transverse mode with a fiber coupler, such that either an MMF or a single-mode fiber (SMF) could serve as the source waveguide~\cite{liu2021fast,sie2020high}, with MMF being a generally less expensive option. After entering the liquid phantom, the light randomly scatters and decorrelates, and a small fraction of which reaches the DMD placed immediately behind the turbid medium. The side of the phantom cuvette facing the DMD is made of microscope slide coverglass. Each square DMD pixel has $13.7\times13.7 \mu \text{m}^2$ area. With $768\times1024$ pixels, the entire DMD panel has a screen size of $10.4\times13.9\text{mm}^2$. We chose to use a DMD to generate the spatiotemporal dynamic scattering patterns first because it is easily configurable: light reaching the quickly flipping pixels decorrelates faster than light that does not, and these pixels are digitally addressible and thus can be changed both spatially and temporally without moving the setup. Second, because it can meet requisite dynamic variation speeds (we run the DMD between 5-10 kHz), which we have selected to correlate with the response of blood flow at tested depths (5-8 mm) ~\cite{liu2021fast}. As the reflected multi-scattered light penetrates on average about $\nicefrac{1}{2}-\nicefrac{2}{3}$ times the \textit{source-detector} distance ($\rho$) deep into the phantom tissue~\cite{patterson1995absorption}, we place 12 multi-speckle detection fibers circularly around the source in the center with $\rho=9.0\text{mm}$. Each multi-speckle detection fiber is a MMF with a $250\mu\text{m}$ core diameter and $0.5$ NA. Quantitative plots of an x-z cross section of the most probable scattered and collected photon trajectories, as well as the expected number of photons per speckle per sampling period, are provided in Supplement Fig. S1(B). We use a modern Monte Carlo simulator called "Multi-Scattering"~\cite{jonsson2020multi} that models anisotropy from spherical scattering centers using a Lorenz-Mie based scattering phase functions. The model has recently been rigorously validated against experimental results as shown in ~\cite{Frantz2021,jonsson2021multi} and can obtain 3D representations of photon paths within the simulated scattering medium. Such results are shown in Fig.\ref{fig:fig2_sche}(C) for the experimental configuration presented in this article, where 12 optical fibers are used for collecting photons, which is the imaging space of our PaDI system. Visualizations of 3D trajectories for detected photon using different numbers of fibers are also provided in Supplement Fig. S2(B). 
Away from the tissue surface, the distal ends of the 12 MMFs are bundled together and imaged onto the SPAD array (PF32, Photon Force, UK) with a magnification $M=\nicefrac{z_2}{z_1}$ using a single lens with an iris diaphragm placed directly adjacent to the lens. As labeled in Fig.\ref{fig:fig2_sche}(B), $r$, $z_1$, and $z_2$ are the radius of the iris diaphragm, the distance between fiber bundle exist and lens, and the distance between lens and SPAD array sensor plane, respectively. To form an image of the fiber bundle on the camera, $z_1$ and $z_2$ satisfy the thin lens equation. In practice, $\nicefrac{r}{z_1}$ is much smaller than the fiber NA that we choose, which determines the NA of the overall speckle imaging system. As illustrated in Supplement Fig. 4(B), the $32\times32$ SPAD array has an overall size
of $1.6\times1.6\text{mm}^2$ with a pixel pitch of $w_p = 50\mu\text{m}$ and an active area that is $\phi=6.95\mu\text{m}$ in diameter. As the magnification is fixed for imaging the light exiting the fiber bundle onto the whole camera, we tune the radius of the iris diaphragm to alter the average speckle size, such that approximately 1 speckle on average is mapped onto each SPAD pixel active area; i.e., we want the speckle size on the sensor plane to match $\phi$. Given that the collected light experiences $\sim440$ scattering events on average (see Supplement Fig.1(C)), the emerging light at the tissue surface is a fully developed speckle pattern with an average speckle size of $\nicefrac{\lambda}{2}$ ~\cite{mecklenbrauker1997wigner} and uniformly distributed phase ~\cite{goodman2015statistical}. Hence, setting  $M\lambda /2\text{NA} =\phi$ gives the desired iris radius $r = \nicefrac{\lambda M}{2\phi z_1}$.

\subsection{Supervised learning for image reconstruction}
\label{subsec::method_imaging}
For our first demonstration of PaDI, we use an artificial neural network to reconstruct images and video of deep temporal dynamics from measured surface speckle intensity autocorrelation curves. As detailed in \textit{Parallelized Diffuse Correlation Imaging} and \textit{Data acquisition and preprocessing} subsections, we collect speckles from 12 distinct surface positions using multimode fibers (MMF), and estimate the intensity autocorrelation for each location. Each intensity autocorrelation curve has 400 sampled time-lags ($1.5\mu$s sampling rate). There are 12 such curves, each computed from the associated SPAD pixels that measure scattered light from the PaDI probe's 12 fiber detectors. A new set of such 12 curves is produced every frame integration time $T_{int}$ (variable between 0.1s and 0.4s). Combining and vectorizing our system's 12 autocorrelation curves gives the neural network input, $\vx\in\R^{4800}$. The output of the neural network is an image  $\vx\in\R^{48\times64}$, with an image pixel size of $220\times220\mu\text{m}^2$. This pixel size is a tunable parameter in our reconstruction model, which we select as smaller than the expected achievable resolution~\cite{durduran2014diffuse,liu2021fast}).  

Figure S5 in the supplement depicts our image reconstruction network. While prior works~\cite{li2018deep,li2018imaging,rahmani2018multimode,lyu2019learning,sun2021scalable,gao2021deep} have used image-to-image translation networks to form images of fixed objects through scattering material, our reconstruction task here is quite different from these alternative networks and thus required us to develop a tailored network architecture. First, the format of our network input is unique (multiple autocorrelations created from non-invasive measurement of second-order temporal statistics of scattered light). Second, the contrast mechanism of our network output is also different - a spatial map of dynamic variation described by speed of change per pixel. Our network mapping problem (multi-autocorrelation inputs into spatial maps of temporal dynamics) is thus in some ways similar to domain transform problems. Therefore, our employed network design is most similar to that introduced by Zhu \textit{et.al.}~\cite{zhu2018image}. Overall, the network is composed of an encoder $f_\theta(\cdot)$ to compress the input into a low-dimensional manifold, and a decoder $g_\theta(\cdot)$ to retrieve the spatial map of temporal dynamics from the embedding. The encoder is composed of three fully-connected layers, with skip connections to allow the error to propagate more easily. All fully-connected layers uses leaky-ReLU activation functions with a slope of 0.1, and the first three fully-connected layers have a dropout rate of $0.05$. After the inputs are embedded into a low-dimensional manifold, the decoder maps the embedding into the 2D reconstruction of dynamics using five transposed convolution layers with stride 2 and padding 1. The network is updated to solve the following problem:

\begin{equation}
    \min\limits_{\theta}\sum_{i=1}^{M}\Big(\mathcal{D}(\vx^i,\hat{\vx}^i)+\mathcal{R}(\hat{\vx}^i) \Big)
\end{equation}

where $\hat{\vx}^i\coloneqq g_{\theta}(f_{\theta}(\vy^i))$ is the output of the network from $i^{th}$ set of measurements $\vy_i$, and $M$ is the total number of training pairs. 

\begin{equation}
    \mathcal{D}(\hat{\vx}) = \frac{1}{2}\|\hat{\vx}^i-\vx^i\|_2^2
\end{equation}

is the data-fidelity term that train the network to find prediction that matches the ground truth, and 

\begin{equation}
    \mathcal{R}(\hat{\vx}) = \lambda\|\hat{\vx}\|_1+\gamma\text{TV}(\hat{\vx}). 
\end{equation}

The $\ell_1$ norm is used to promote sparsity of the reconstruction, and $\text{TV}(\cdot)$ is the isotropic total variation penalty that makes the reconstruction piecewise constant. These regularizations have been successfully applied to improve diffuse optics imaging reconstructions~\cite{correia2011split,xu2019improving}. $\lambda$ and $\gamma$ are hyperparameters empirically chosen to be $0.02$ and $0.1$, respectively, to balance the data fidelity and image prior knowledge. As we have a small dataset for training, we do not divide the data further to create a validation dataset. Instead, we apply early stop to avoid overfitting. The networks for all tasks used Xavier initialization~\cite{glorot2010understanding} and trained for 2000 epochs using the Adam optimizer~\cite{kingma2014adam} with a $8\times10^{-4}$ learning rate and $256$ batch size.

\subsection{Experimental validation with digital and bio-inspired phantoms}
\label{sec:result}
\begin{figure}[t!]
\begin{center}
\includegraphics[width=8.0cm]{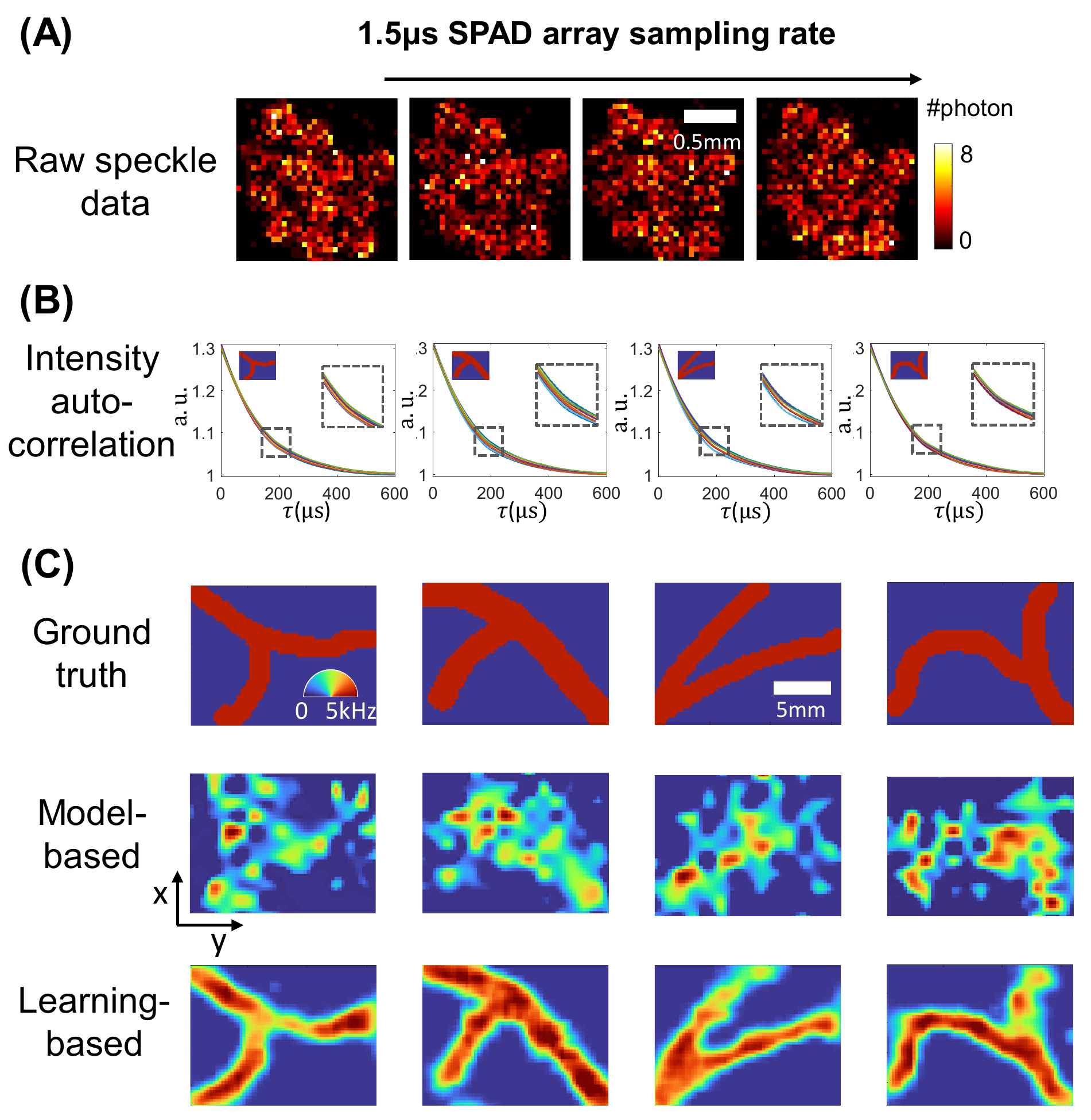}
\end{center}
\caption{PaDI Measurements and reconstructions of phantom vasculature patterns located 5mm beneath a tissue-like decorrelating turbid volume. (A) Recorded raw SPAD array speckle intensity (colorbar: photons detected per pixel). (B) Processed intensity auto-correlations using $T_{int}=0.4$s where x-axis is time-lag $\tau$. Each plot labeled with ground truth of dynamic scattering image on the top-left, with zoom-ins showing curve regions most sensitive to spatially varying decorrelation. (C) Ground truth dynamic scattering object 5 mm beneath tissue phantom with PaDI reconstructions using a model-based method (for comparison) and proposed learning-based method. All figures in (C) share same color wheel (dynamic scatter fluctuation rate), scale bar, and x-y coordinates}
\label{fig:fig3_vas}
\end{figure}
We validated our learning-based image reconstruction method with four unique experiments that each utilized a unique training data set. First, since detecting deep-tissue blood flow is a primary aim of PaDI system development, we studied the ability of our network to image vessel-like structures using 1428 \emph{vasculature} patterns extracted from biomedical image data~\cite{yao2014photoacoustic} (1190 for training, 238 for testing). Image data was rescaled to an appropriate size ($10.4\times13.9\text{mm}^2$) and displayed as a dynamic pattern with a 5 kHz variation rate. By comparing PaDI and standard inverse diffusion model-based reconstructions (as detailed in supplement Section 3), we highlight significant improvement. Second, we tested the generalizability of PaDI by training the network with objects drawn from one type of dataset, and testing the network with objects drawn from a second distinct dataset type (i.e., from a different distribution). For this generalizability experiment, we trained with 1280 hand-written \emph{letters} from the EMNIST dataset and assessed reconstruction accuracy used 128 \emph{digits} from the MNIST dataset during algorithm testing. Third, we explored the potential of our method to jointly image both temporal and spatially varying dynamic potentials by using PaDI to image objects of different sizes and unique fluctuation rates ($5$ kHz and $10$ kHz). Finally, we further tested system generalizability by acquiring PaDI data from a completely unique phantom tissue arrangement, containing 3 mm diameter phantom vessels buried 5 mm beneath scattering material, through which we flowed liquid at variable speeds. We then applied a DMD-trained image formation model, trained with 1046 patterns containing two tube-shape objected demonstrated in Fig.\ref{fig:fig5_syringPump}(A), to spatio-temporally resolve buried capillary flow dynamics, highlighting the flexibility of both the imaging hardware and post-processing software.

 \begin{figure}[t!]
\begin{center}
\includegraphics[width=15.0cm]{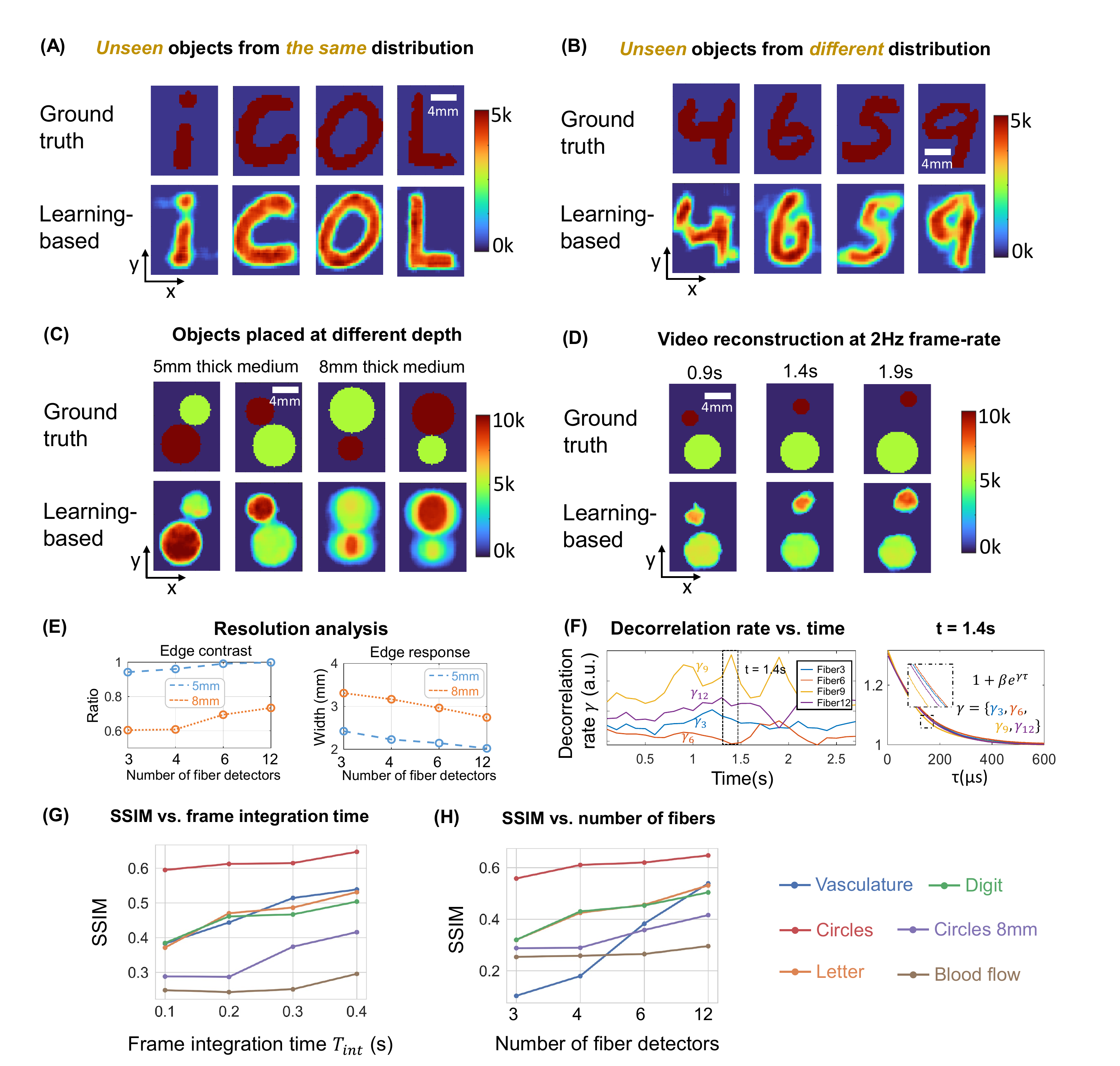}
\end{center}
\caption{PaDI reconstructions of spatiotemporal dynamics for various patterns and decorrelation speeds hidden beneath 5mm-8mm thick turbid volume. (A) Reconstructions of letter-shaped dynamic scatter patterns hidden underneath 5mm turbid volume, sampled from a distribution that matches training data distribution. (B) Reconstructions of digit-shaped  dynamic scatter patterns hidden underneath 5mm turbid volume, drawn from a different distribution as compared to training data distribution. (C) Reconstructions of objects at varying dynamic scattering rate hidden beneath 5mm and 8mm-thick turbid volume, along with (E) Resolution analyses for different depths using different number of fiber detectors. (D) A few reconstruction frames from a video taken over 3s. (F) plots four of decorrelation rates change in time. A set of autocorrelation curves from these four-fiber detection at 1.4s is presented on the right.  (G) Plots of average SSIM between ground-truth and reconstructed speed maps as a function of frame integration time $T_{int}$ for various tested datasets. (H) Plots of average SSIM between ground-truth and reconstructed speed maps as a function of number of detection fibers used for image formation. (G) and (H) share the same legend listed at the bottom of the figure. Imaging datasets are described in the \textit{Experimental validation} sub-section.}
\label{fig:fig4_digitresult}
\end{figure}
Figure \ref{fig:fig3_vas}(A) shows a few representative raw SPAD array measurements ($1.5\mu$s exposure time). 12 circular spots in the raw frame are roughly discernible. Each spot contains photon count statistics of scattered light collected from one of 12 different locations on the tissue phantom surface and delivered to the array via MMF. Figure \ref{fig:fig3_vas}(B) plots the intensity autocorrelation curves for each of the 12 unique SPAD array regions (i.e., each unique location on the tissue phantom surface). These curves are averages computed over space (all SPAD measurements per fiber) and time (a frame integration time here of $0.4$s). The dynamic scattering patterns used to generate each set of auto-correlation curves are labeled on the upper right corner of each plot, and the regions most sensitive to the perturbations are enlarged. 
 
 The first row of Fig.\ref{fig:fig3_vas}(C) displays several examples of dynamic patterns from the \emph{vasculature} dataset produced in our phantom setup beneath 5 mm of turbid decorrelating media. The second and third rows show PaDI reconstructions for these patterns using our proposed learning-based method and a regularized model-based reconstruction method, for comparison. Details regarding the model-based reconstruction method can be found in Supplementary Section 3. Due to the ill-posed nature of the inverse problem and model-experiment mismatch, model-based reconstruction results are less spatially informative compared to our proposed learning-based method, even when strong structural image priors are used. We observe some marginal artifacts in reconstructions using the proposed learning-based method, where the reconstructed edge values are typically lower than the ground-truth, as the high frequency on the edge is harder to reconstruct. 
 While Fig. \ref{fig:fig4_digitresult}(A) shows the dynamic scattering potential reconstructions for \emph{unseen} objects drawn from a distribution that matches the training dataset, Fig. \ref{fig:fig4_digitresult}(B) shows dynamic scattering reconstructions for \emph{unseen} objects drawn from a \emph{different} distribution as compared to the training dataset. These results suggest that the trained network has the generalizability to predict unseen dynamic scattering objects that have limited correlation with expectation. At the same time, we also observe that the reconstructions for the objects drawn from a different distribution are of less sharp than reconstructions for objects drawn from the same distribution as the training set, even though the average structural similarity index measure(SSIM)~\cite{wang2004image} values between the two testing datasets are comparable, as shown in Fig.\ref{fig:fig4_digitresult}(G)-(H). 
 
 Next, we tested the ability of PaDI to resolve decorrelation speed maps that vary as a function of space and at different phantom tissue depths. PaDI reconstructions for two variable-speed perturbations under both 5mm and 8mm of turbid medium are in Fig.\ref{fig:fig4_digitresult}(C). In this experiment, 1280 and 108 patterns of variable speed and shape were utilized for training and testing, respectively. First, we observe that PaDI can spatially resolve features while still maintaining an accurate measure of unique decorrelation speeds. When structures with different decorrelation speeds  begin to spatially overlap, the associated reconstructed speed values close to the overlap boundary are either lifted or lowered towards that of the neighboring structure. This is expected, as the detected light travelling through the "banana-shaped" light path contains information integrated over a finite-sized sensitivity region that will effectively limit the spatial resolution of the associated speed map reconstruction. Moreover, we also observed that PaDI reconstructions of dynamics hidden beneath a thicker 8mm scattering medium are less accurate than those for dynamics beneath a 5mm scattering medium. A resolution analysis based on the contrast in the edge regions of the circles, and the width of $10\%-90\%$ edge response is also provided in Fig.\ref{fig:fig4_digitresult}(E). Speckle fluctuations sampled by our current configuration on the phantom tissue surface are less sensitive to decorrelation events occurring within deeper region. Creating a PaDI probe with larger source-detector separations can help address this challenge, as detailed in the discussion section. Further, we collect continuous data for 3s, where the dynamic patterns hidden underneath present for 0.3s, and change every 0.5s. We show reconstructions of a few frames at Fig.\ref{fig:fig4_digitresult}(D). Fig.\ref{fig:fig4_digitresult}(F) plots four of decorrelation rates change in time. The decorrelation rates are extracted by fitting each autocorrelation curves with with $1+\beta e^{\gamma \tau}$, where $\tau$ is delay-time and $\gamma$ is the decorrelation rates. These autocorrelation curves are used to generate reconstructions in Fig.\ref{fig:fig4_digitresult}(D). 3 seconds continuous measurements are taken, and the curves are estimated using 0.3s integration window and $66.7\%$ overlap between sliding windows. A set of autocorrelation curves from these four-fiber detection at 1.4s is presented on the right.  
 \begin{figure}[t!]
\begin{center}
\includegraphics[width=14cm]{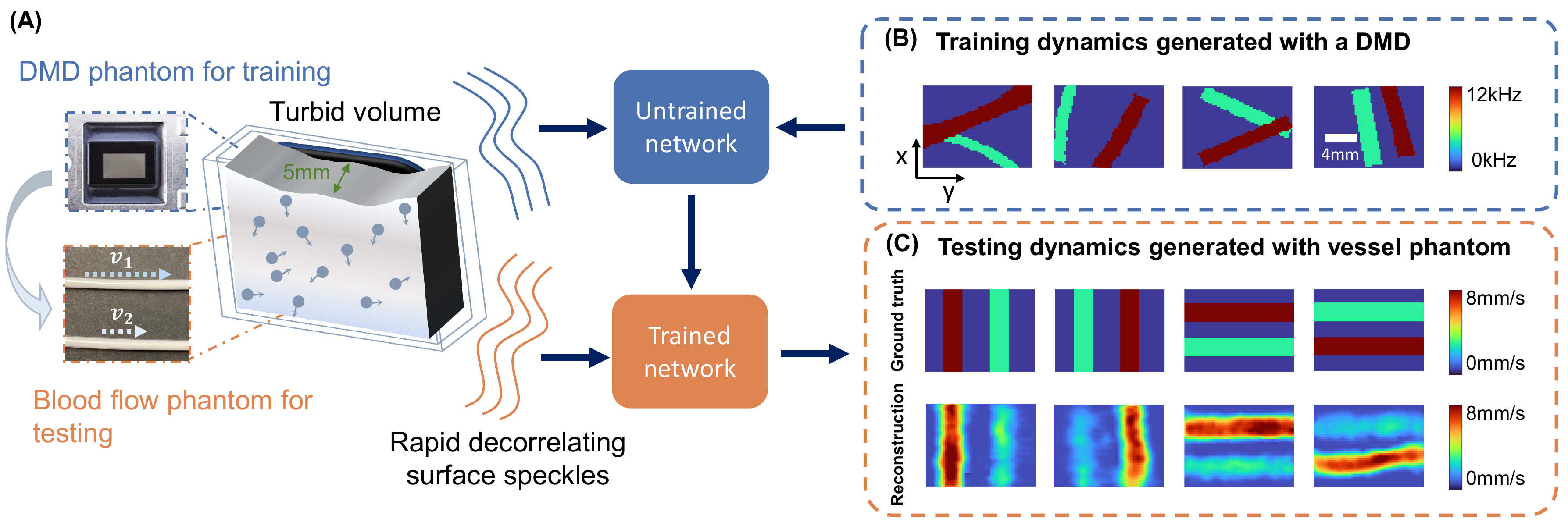}
\end{center}
\caption{(A) Illustration of deep tissue phantom capillary flow experiment. PaDI network is first trained on synthetic data generated by DMD phantom, then applied to reconstruct images from separate capillary flow phantom setup. (B) Examples of dynamic scattering patterns used for training, generated at up to 12 kHz on DMD phantom. (C) Representative images reconstructed with proposed learning-based method, along with ground truth. Dynamics are generated with two capillary tubes buried beneath a 5 mm scattering volume exhibiting variable-speed liquid flow.}
\label{fig:fig5_syringPump}
\end{figure}
We additionally conducted an experiment to study how our model, trained with data generated on our digital phantom, can reconstruct images of the dynamic scattering introduced by more biologically realistic contrast mechanisms. Non-invasive imaging of deep blood flow dynamics, such as hemodynamics within the human brain, is an important application for diffuse optical correlation-based measurements. Accordingly, we modeled deep hemodynamic flow by placing two capillary tubes (3mm diameter) directly beneath a dynamic scattering volume (same optical properties: $\mu_a=0.01$ mm$^{-1}$, $\mu_s'=0.7$ mm$^{-1}$ ) flowing at two different speeds (2.7mm/s and 8.0mm/s) via syringe pump injection. After training an image formation model with PaDI data captured on our DMD-based phantom (630 maps of randomly oriented tube-like objects varying at 4-12kHz, see Fig.~\ref{fig:fig5_syringPump}(A)), we acquired PaDI data from this unique capillary flow phantom and applied the DMD phantom-trained model to produce images as shown in Figure \ref{fig:fig5_syringPump}(B). Here, we observe reconstructed image measurements of relative flow speed with spatial and temporal structures that match ground truth, pointing towards a system that can potentially image dynamic scattering beneath tissue \textit{in vivo} using learning-based reconstruction methods trained with more easily accessible synthetic data

Finally, we assessed experimental PaDI performance as a function of detection speed and number of spatial measurement points using the structural similarity index measure (SSIM) metric~\cite{wang2004image}, as more than one contrast mechanism (DMD, fluid dynamics) used in different experiments. Fig.\ref{fig:fig4_digitresult}(G-H) plot average SSIM as a function of frame integration time $T_{int}$ and as a function of number of surface detectors $P$ for all datasets above. Fig. \ref{fig:fig4_digitresult}(G)'s data used all 12 unique phantom surface locations for its reconstructions. \ref{fig:fig4_digitresult}(H)'s data used a $0.4$s frame integration time. From these plots, it is clear that a longer frame integration time improves reconstruction performance, at the expense of a proportionally decreased PaDI frame rate. In addition, collecting speckle dynamics from more surface locations improves the reconstruction results, as expected. This is not only because the imaging (photon-sensitive) region of the 12-fiber system is larger than that using fewer fibers, but also because the overlap between banana-shaped photon paths from adjacent fiber detectors (e.g., see Supplement Fig. S2) provides redundant data that is beneficial to accurate image formation.

\subsection{Discussion}
In summary, we have developed a new parallelized speckle sensing method that can spatially resolve maps of decorrelation dynamics that occur beneath multiple millimeters of tissue-like scattering media. Our approach utilizes diffuse correlation principles to sample speckle fluctuations from different locations along a scattering medium's surface at high speed. Unlike prior work, our system records all such measurements in parallel to reconstruct transient speed maps at multi-hertz video frame rate, and uses a novel machine learning approach for this reconstruction task that outperforms standard model-based solvers.  

While we demonstrated that PaDI can rapidly image dynamic events occurring under a decorrelating tissue phantom, several potential improvements can be made to ensure effective translation into \textit{in vivo} use. First, as shown in the raw speckle data from the SPAD array, the fiber bundle we use was not optimized to maximize the speckle detection efficiency - our fiber bundle array did not map surface speckle to all SPADs within the array. Future work will endeavor to utilize a custom-designed fiber bundle that provides better array coverage. We note that detection efficiency was further reduced in our phantom setup by the cover glass surfaces on both sides of the cuvette holding the liquid tissue phantom, both via reflection and by enforcing a finite standoff distance from the phantom for the fiber probe, which decreased light collection efficiency. This can be resolved in the future using a more suitable material ~\cite{wabnitz2014performance}, which we expect to further improve the sensitivity of our PaDI system. In addition, we used a DMD in this work to generate simulated deep-tissue dynamics because it provided an easily reconfigurable means to assess performance for a variety of decorrelating structures. The use of a DMD restricted the total lateral dimension of the phantom tissue and hidden structure that we were able to probe, which additionally prevented us from being able to investigate larger source-detector separations that are well-known to improve detection accuracy for deeper dynamics. Based upon the findings in this work, a tissue phantom with embedded vessel phantoms containing flowing liquid can be designed to provide additional verification of PaDI imaging performance at greater depths~\cite{kleiser2018comparison}. Recently developed time-of-flight~\cite{sutin2016time,kholiqov2020time} methods also enhance signal from greater depths and can be considered as additional avenues through which PaDI can be improved. 

In the future, we also plan to study how our system can jointly image blood flow at variable oxygenation levels. By adding an isosbestic wavelength to the current system, we can potentially spatially resolve blood flow speed as a function of oxygen level. On the computational side, one of the problems of using classic supervised deep learning methods as a maximum likelihood estimator is reconstruction reliability concern. One expensive solution is to expand the training set to include large amount of objects. In this data-rich scenario, meta-learning approaches can also be considered, where part of the network weights are allowed to be changed depending on different imaging setups~\cite{snell2017prototypical}. In a resource limited situation, however, an alternative strategy might  assess the reliability by predicting the uncertainty along with the reconstruction using approximate deep Bayesian inference~\cite{mandt2017stochastic}. These additional investigations will aid with the eventual translation of PaDI into a practical and reliable tool for recording video of deep-tissue blood flow in \emph{in vivo} subjects in the future.  

\section{Experimental Section}
\label{sec:method}
\subsection{Data acquisition and preprocessing}
\label{sec::dataprepros}
We use the SPAD array's $1024$ ($32\times32$) independent single SPADs to count photons arriving at each pixel with a frame rate of $666$kHz and a bit depth of 4. This is equivalent to an exposure time of $T_s=1.5\mu\text{s}$. To extract the temporal statistics from measurements of randomly fluctuating surface speckle at $666$kHz, we then compute a temporal autocorrelation on a per-SPAD basis. Although we note that there are a number of strategies available to compute such temporal statistics across a SPAD array (e.g., joint processing across pixels, examining higher-order statistics, or more advanced autocorrelation inference methods~\cite{lemieux1999investigating,jazani2019alternative}), we have selected the per-pixel method here as it is well-established~\cite{johansson2019multipixel,sie2020high,liu2021fast}. We compute our temporal autocorrelations across ``frame integration time" of typically $T_{int}=0.4$s, which yields $N=\nicefrac{T_{int}}{T_s}$ frames per autocorrelation measurement. Rather than using a physical correlator module, we record the time-resolved photon stream as a $1024\times N$ array and compute the autocorrelations in software, where typically $N=266$k. We also explore the effect of using a shorter $T_{int}$ and fewer SPADs per measurement and compare the results in Fig.\ref{fig:fig4_digitresult}(G-H). 

As illustrated in Supplement Fig.4, we compute the normalized temporal intensity autocorrelation ~\cite{durduran2014diffuse} of each pixel as, 

\begin{equation}
    g_2^{p,q}(\tau) = \frac{\langle I^{p,q}(t)I^{p,q}(t+\tau)\rangle_{T_{int}}}{\langle I^{p,q}(t){\rangle}_{T_{int}}^2},
\label{eq:correlation}
\end{equation}

where $I^{p,q}(t)$ is the photon count detected by the $q$-th SPAD for $p$-th fiber at time $t$; $\tau$ is time-lag (or delay or correlation time), and ${\langle}\cdot{\rangle}_{T_{int}}$ denotes time average estimated by integrating over $T_{int}$. After calculating $g_2^{p,q}(\tau)$ for each single SPAD, we then obtain an average, noise-reduced curve by averaging curves that are produced by the $Q_p$ unique SPADs that detect light emitted by the same MMF detection fiber: 

\begin{equation}
    \overline{g}_2^{p}(\tau)=\frac{1}{Q_p}\sum_{q=1}^{Q_p}g_2^{p,q}
\label{eq:average_corr}
\end{equation}

for the $p^{th}$ MMF fiber, where we used a total of 12 MMF. A straightforward calibration procedure allows us to identify the $Q_p$ SPADs within the array that receives light from the $p$th MMF, which we save as a look-up table. We next compile the $g_2^{p}(\tau)$ from each fiber into a set of 12 average intensity autocorrelation curves per frame, with the aim of reconstructing the spatiotemporal scattering structure hidden beneath the decorrelating phantom. An example set of intensity autocorrelation curves is in Fig.~\ref{fig:fig1_flow}(B). The maximum lag or delay time $\tau_{\text{max}}$ is selected at $600\mu\text{s}$, as the values of the intensity autocorrelation start approaching $1$ asymptotically.

\section*{Acknowledgements}
Research reported in this publication was supported by the National Institute of Neurological Disorders and Stroke of the National Institutes of Health under award number RF1NS113287, as well as the Duke-Coulter Translational Partnership. The authors also want to thank Kernel Inc. for their generous support. W.L. acknowledges the support from the China Scholarship Council. R.H. acknowledges support from a Hartwell Foundation Individual Biomedical Researcher Award, and Air Force Office of Scientific Research under award number FA9550. In addition, the authors would like to express our great appreciation to Dr. Haowen Ruan for editing and inspirational discussion.

\section*{Data availability}
The data support the findings of this study are available from the corresponding author through collaborative investigations and upon reasonable request.

\section*{Author contributions statement}
S. X., X. Y., W. L. R. Q. and P. C. K. constructed the hardware setup. S. X., W. L. and J. J. designed the software. S. X, K. Z, L. K, E. B. and R. H wrote the manuscript. H. W., Q. D., E. B. and R. H. supervised the project. 
\section*{Conflict of interest}
S.X. and R.H. have submitted a patent application for this work, assigned to Duke University
\section*{Supplement}
\href{https://waltzina.github.io/files/PaDI_supp.pdf}{Supplement material}.

\medskip

%
\bibliographystyle{unsrt}  
\bibliography{sample}  

\begin{thebibliography}{10}

\bibitem{ntziachristos2010going}
Vasilis Ntziachristos.
\newblock Going deeper than microscopy: the optical imaging frontier in
  biology.
\newblock {\em Nature methods}, 7(8):603--614, 2010.

\bibitem{horton2013vivo}
Nicholas~G Horton, Ke~Wang, Demirhan Kobat, Catharine~G Clark, Frank~W Wise,
  Chris~B Schaffer, and Chris Xu.
\newblock In vivo three-photon microscopy of subcortical structures within an
  intact mouse brain.
\newblock {\em Nature photonics}, 7(3):205--209, 2013.

\bibitem{horstmeyer2015guidestar}
Roarke Horstmeyer, Haowen Ruan, and Changhuei Yang.
\newblock Guidestar-assisted wavefront-shaping methods for focusing light into
  biological tissue.
\newblock {\em Nature photonics}, 9(9):563--571, 2015.

\bibitem{lyons2019computational}
Ashley Lyons, Francesco Tonolini, Alessandro Boccolini, Audrey Repetti, Robert
  Henderson, Yves Wiaux, and Daniele Faccio.
\newblock Computational time-of-flight diffuse optical tomography.
\newblock {\em Nature Photonics}, 13(8):575--579, 2019.

\bibitem{lindell2020three}
David~B Lindell and Gordon Wetzstein.
\newblock Three-dimensional imaging through scattering media based on confocal
  diffuse tomography.
\newblock {\em Nature communications}, 11(1):1--8, 2020.

\bibitem{wang2012photoacoustic}
Lihong~V Wang and Song Hu.
\newblock Photoacoustic tomography: in vivo imaging from organelles to organs.
\newblock {\em science}, 335(6075):1458--1462, 2012.

\bibitem{jang2015relation}
Mooseok Jang, Haowen Ruan, Ivo~M Vellekoop, Benjamin Judkewitz, Euiheon Chung,
  and Changhuei Yang.
\newblock Relation between speckle decorrelation and optical phase conjugation
  (opc)-based turbidity suppression through dynamic scattering media: a study
  on in vivo mouse skin.
\newblock {\em Biomedical optics express}, 6(1):72--85, 2015.

\bibitem{wang2015focusing}
Daifa Wang, Edward~Haojiang Zhou, Joshua Brake, Haowen Ruan, Mooseok Jang, and
  Changhuei Yang.
\newblock Focusing through dynamic tissue with millisecond digital optical
  phase conjugation.
\newblock {\em Optica}, 2(8):728--735, 2015.

\bibitem{tzang2019wavefront}
Omer Tzang, Eyal Niv, Sakshi Singh, Simon Labouesse, Greg Myatt, and Rafael
  Piestun.
\newblock Wavefront shaping in complex media with a 350 khz modulator via a
  1d-to-2d transform.
\newblock {\em Nature Photonics}, 13(11):788--793, 2019.

\bibitem{ruan2020fluorescence}
Haowen Ruan, Yan Liu, Jian Xu, Yujia Huang, and Changhuei Yang.
\newblock Fluorescence imaging through dynamic scattering media with
  speckle-encoded ultrasound-modulated light correlation.
\newblock {\em Nature Photonics}, 14(8):511--516, 2020.

\bibitem{gigan2021roadmap}
Sylvain Gigan, Ori Katz, Hilton~B de~Aguiar, Esben~Ravn Andresen, Alexandre
  Aubry, Jacopo Bertolotti, Emmanuel Bossy, Dorian Bouchet, Joshua Brake,
  Sophie Brasselet, et~al.
\newblock Roadmap on wavefront shaping and deep imaging in complex media.
\newblock {\em arXiv preprint arXiv:2111.14908}, 2021.

\bibitem{korolevich2000experimental}
Alexander~N Korolevich and Igor~V Meglinsky.
\newblock Experimental study of the potential use of diffusing wave
  spectroscopy to investigate the structural characteristics of blood under
  multiple scattering.
\newblock {\em Bioelectrochemistry}, 52(2):223--227, 2000.

\bibitem{dunn2001dynamic}
Andrew~K Dunn, Hayrunnisa Bolay, Michael~A Moskowitz, and David~A Boas.
\newblock Dynamic imaging of cerebral blood flow using laser speckle.
\newblock {\em Journal of Cerebral Blood Flow \& Metabolism}, 21(3):195--201,
  2001.

\bibitem{culver2003diffuse}
Joseph~P Culver, Turgut Durduran, Daisuke Furuya, Cecil Cheung, Joel~H
  Greenberg, and AG~Yodh.
\newblock Diffuse optical tomography of cerebral blood flow, oxygenation, and
  metabolism in rat during focal ischemia.
\newblock {\em Journal of cerebral blood flow \& metabolism}, 23(8):911--924,
  2003.

\bibitem{kim2003intracellular}
Sally~A Kim and Petra Schwille.
\newblock Intracellular applications of fluorescence correlation spectroscopy:
  prospects for neuroscience.
\newblock {\em Current opinion in neurobiology}, 13(5):583--590, 2003.

\bibitem{wang2007three}
Ruikang~K Wang, Steven~L Jacques, Zhenhe Ma, Sawan Hurst, Stephen~R Hanson, and
  Andras Gruber.
\newblock Three dimensional optical angiography.
\newblock {\em Optics express}, 15(7):4083--4097, 2007.

\bibitem{yao2010vivo}
Junjie Yao, Konstantin~I Maslov, Yunfei Shi, Larry~A Taber, and Lihong~V Wang.
\newblock In vivo photoacoustic imaging of transverse blood flow by using
  doppler broadening of bandwidth.
\newblock {\em Optics letters}, 35(9):1419--1421, 2010.

\bibitem{durduran2014diffuse}
Turgut Durduran and Arjun~G Yodh.
\newblock Diffuse correlation spectroscopy for non-invasive, micro-vascular
  cerebral blood flow measurement.
\newblock {\em Neuroimage}, 85:51--63, 2014.

\bibitem{buckley2014diffuse}
Erin~M Buckley, Ashwin~B Parthasarathy, P~Ellen Grant, Arjun~G Yodh, and
  Maria~Angela Franceschini.
\newblock Diffuse correlation spectroscopy for measurement of cerebral blood
  flow: future prospects.
\newblock {\em Neurophotonics}, 1(1):011009, 2014.

\bibitem{boas1995scattering}
David~A Boas, LE~Campbell, and Arjun~G Yodh.
\newblock Scattering and imaging with diffusing temporal field correlations.
\newblock {\em Physical review letters}, 75(9):1855, 1995.

\bibitem{bruschini2019single}
Claudio Bruschini, Harald Homulle, Ivan~Michel Antolovic, Samuel Burri, and
  Edoardo Charbon.
\newblock Single-photon avalanche diode imagers in biophotonics: review and
  outlook.
\newblock {\em Light: Science \& Applications}, 8(1):1--28, 2019.

\bibitem{morimoto2020megapixel}
Kazuhiro Morimoto, Andrei Ardelean, Ming-Lo Wu, Arin~Can Ulku, Ivan~Michel
  Antolovic, Claudio Bruschini, and Edoardo Charbon.
\newblock Megapixel time-gated spad image sensor for 2d and 3d imaging
  applications.
\newblock {\em Optica}, 7(4):346--354, 2020.

\bibitem{canon2021mega}
Canon successfully develops the world’s first 1-megapixel spad sensor.
\newblock {\em https://global.canon/en/technology/spad-sensor-2021.html},
  accessed: 06.14.2021.

\bibitem{zickus2020wide}
Vytautas Zickus, Ming~Lo Wu, Kazuhiro Morimoto, Valentin Kapitany, Areeba
  Farima, Alex Turpin, Robert Insall, Jamie Whitelaw, Laura Machesky, Claudio
  Bruschini, et~al.
\newblock Wide-field fluorescence lifetime imaging microscopy with a high-speed
  mega-pixel spad camera.
\newblock {\em bioRxiv}, 2020.

\bibitem{buttafava2020spad}
Mauro Buttafava, Federica Villa, Marco Castello, Giorgio Tortarolo, Enrico
  Conca, Mirko Sanzaro, Simonluca Piazza, Paolo Bianchini, Alberto Diaspro,
  Franco Zappa, et~al.
\newblock Spad-based asynchronous-readout array detectors for image-scanning
  microscopy.
\newblock {\em Optica}, 7(7):755--765, 2020.

\bibitem{slenders2021confocal}
Eli Slenders, Marco Castello, Mauro Buttafava, Federica Villa, Alberto Tosi,
  Luca Lanzan{\`o}, Sami~Valtteri Koho, and Giuseppe Vicidomini.
\newblock Confocal-based fluorescence fluctuation spectroscopy with a spad
  array detector.
\newblock {\em Light: Science \& Applications}, 10(1):1--12, 2021.

\bibitem{Xi2021quan}
Xi~Yang, Pavan Chandra~Konda, Shiqi Xu, Liheng Bian, and Roarke Horstmeyer.
\newblock Quantized fourier ptychography with binary images from spad cameras.
\newblock {\em Photonics Research, In Press}, 2021.

\bibitem{gyongy2020high}
Istvan Gyongy, Sam~W Hutchings, Abderrahim Halimi, Max Tyler, Susan Chan, Feng
  Zhu, Stephen McLaughlin, Robert~K Henderson, and Jonathan Leach.
\newblock High-speed 3d sensing via hybrid-mode imaging and guided upsampling.
\newblock {\em Optica}, 7(10):1253--1260, 2020.

\bibitem{gariepy2016detection}
Genevieve Gariepy, Francesco Tonolini, Robert Henderson, Jonathan Leach, and
  Daniele Faccio.
\newblock Detection and tracking of moving objects hidden from view.
\newblock {\em Nature Photonics}, 10(1):23, 2016.

\bibitem{satat2016all}
Guy Satat, Barmak Heshmat, Dan Raviv, and Ramesh Raskar.
\newblock All photons imaging through volumetric scattering.
\newblock {\em Scientific reports}, 6(1):1--8, 2016.

\bibitem{zhou2006diffuse}
Chao Zhou, Guoqiang Yu, Daisuke Furuya, Joel~H Greenberg, Arjun~G Yodh, and
  Turgut Durduran.
\newblock Diffuse optical correlation tomography of cerebral blood flow during
  cortical spreading depression in rat brain.
\newblock {\em Optics express}, 14(3):1125--1144, 2006.

\bibitem{han2015non}
Songfeng Han, Michael~D Hoffman, Ashley~R Proctor, Joseph~B Vella, Emmanuel~A
  Mannoh, Nathaniel~E Barber, Hyun~Jin Kim, Ki~Won Jung, Danielle~SW Benoit,
  and Regine Choe.
\newblock Non-invasive monitoring of temporal and spatial blood flow during
  bone graft healing using diffuse correlation spectroscopy.
\newblock {\em PLoS One}, 10(12):e0143891, 2015.

\bibitem{he2015noncontact}
Lian He, Yu~Lin, Chong Huang, Daniel Irwin, Margaret~M Szabunio, and Guoqiang
  Yu.
\newblock Noncontact diffuse correlation tomography of human breast tumor.
\newblock {\em Journal of biomedical optics}, 20(8):086003, 2015.

\bibitem{johansson2019multipixel}
Johannes~D Johansson, Davide Portaluppi, Mauro Buttafava, and Federica Villa.
\newblock A multipixel diffuse correlation spectroscopy system based on a
  single photon avalanche diode array.
\newblock {\em Journal of biophotonics}, 12(11):e201900091, 2019.

\bibitem{liu2020classifying}
Wenhui Liu, Shiqi Xu, Ruobing Qian, Pavan~Chanda Konda, and Roarke Horstmeyer.
\newblock Classifying decorrelation events hidden beneath scattering media via
  spad array detection.
\newblock In {\em Computational Optical Sensing and Imaging}, pages CTu5A--3.
  Optical Society of America, 2020.

\bibitem{liu2021fast}
Wenhui Liu, Ruobing Qian, Shiqi Xu, Pavan Chandra~Konda, Joakim J{\"o}nsson,
  Mark Harfouche, Dawid Borycki, Colin Cooke, Edouard Berrocal, Qionghai Dai,
  et~al.
\newblock Fast and sensitive diffuse correlation spectroscopy with highly
  parallelized single photon detection.
\newblock {\em APL Photonics}, 6(2):026106, 2021.

\bibitem{sie2020high}
Edbert~J Sie, Hui Chen, E-Fann Saung, Ryan Catoen, Tobias Tiecke, Mark~A
  Chevillet, and Francesco Marsili.
\newblock High-sensitivity multispeckle diffuse correlation spectroscopy.
\newblock {\em Neurophotonics}, 7(3):035010, 2020.

\bibitem{xu2021rapid}
Shiqi Xu, Xi~Yang, Pavan~Chanda Konda, and Roarke Horstmeyer.
\newblock Rapid imaging of deep-tissue motion with parallelized diffuse
  correlation spectroscopy.
\newblock In {\em Optics and the Brain}, pages BTh1B--3. Optical Society of
  America, 2021.

\bibitem{zhou2021functional}
Wenjun Zhou, Oybek Kholiqov, Jun Zhu, Mingjun Zhao, Lara~L Zimmermann, Ryan~M
  Martin, Bruce~G Lyeth, and Vivek~J Srinivasan.
\newblock Functional interferometric diffusing wave spectroscopy of the human
  brain.
\newblock {\em Science Advances}, 7(20):eabe0150, 2021.

\bibitem{xu2021diffusing}
Jian Xu, Ali~K Jahromi, and Changhuei Yang.
\newblock Diffusing wave spectroscopy: A unified treatment on temporal sampling
  and speckle ensemble methods.
\newblock {\em APL Photonics}, 6(1):016105, 2021.

\bibitem{xu2022speckle}
Shiqi Xu, Xi~Yang, Joakim J{\"o}nsson, Hansori Chang, and Roarke Horstmeyer.
\newblock Speckle contrast diffuse correlation spectroscopy with parallelized
  single photon detection.
\newblock In {\em Optics and the Brain}, pages BTu2C--3. Optica Publishing
  Group, 2022.

\bibitem{xu2022transient}
Shiqi Xu, Wenhui Liu, Xi~Yang, Joakim J{\"o}nsson, Ruobing Qian, Paul McKee,
  Kanghyun Kim, Pavan~Chandra Konda, Kevin~C Zhou, Lucas Krei{\ss}, et~al.
\newblock Transient motion classification through turbid volumes via
  parallelized single-photon detection and deep contrastive embedding.
\newblock {\em arXiv preprint arXiv:2204.01733}, 2022.

\bibitem{horisaki2016learning}
Ryoichi Horisaki, Ryosuke Takagi, and Jun Tanida.
\newblock Learning-based imaging through scattering media.
\newblock {\em Optics express}, 24(13):13738--13743, 2016.

\bibitem{li2018deep}
Yunzhe Li, Yujia Xue, and Lei Tian.
\newblock Deep speckle correlation: a deep learning approach toward scalable
  imaging through scattering media.
\newblock {\em Optica}, 5(10):1181--1190, 2018.

\bibitem{li2018imaging}
Shuai Li, Mo~Deng, Justin Lee, Ayan Sinha, and George Barbastathis.
\newblock Imaging through glass diffusers using densely connected convolutional
  networks.
\newblock {\em Optica}, 5(7):803--813, 2018.

\bibitem{rahmani2018multimode}
Babak Rahmani, Damien Loterie, Georgia Konstantinou, Demetri Psaltis, and
  Christophe Moser.
\newblock Multimode optical fiber transmission with a deep learning network.
\newblock {\em Light: Science \& Applications}, 7(1):1--11, 2018.

\bibitem{lyu2019learning}
Meng Lyu, Hao Wang, Guowei Li, Shanshan Zheng, and Guohai Situ.
\newblock Learning-based lensless imaging through optically thick scattering
  media.
\newblock {\em Advanced Photonics}, 1(3):036002, 2019.

\bibitem{sun2021scalable}
Yiwei Sun, Xiaoyan Wu, Yuanyi Zheng, Jianping Fan, and Guihua Zeng.
\newblock Scalable non-invasive imaging through dynamic scattering media at low
  photon flux.
\newblock {\em Optics and Lasers in Engineering}, 144:106641, 2021.

\bibitem{gao2021deep}
Ya~Gao, Wenyi Xu, Yiming Chen, Weiya Xie, and Qian Cheng.
\newblock Deep learning-based photoacoustic imaging of vascular network through
  thick porous media.
\newblock {\em arXiv preprint arXiv:2103.13964}, 2021.

\bibitem{boas2016establishing}
David~A Boas, Sava Sakad{\v{z}}i{\'c}, Juliette~J Selb, Parisa Farzam,
  Maria~Angela Franceschini, and Stefan~A Carp.
\newblock Establishing the diffuse correlation spectroscopy signal relationship
  with blood flow.
\newblock {\em Neurophotonics}, 3(3):031412, 2016.

\bibitem{durduran2010diffuse}
Turgut Durduran, Regine Choe, Wesley~B Baker, and Arjun~G Yodh.
\newblock Diffuse optics for tissue monitoring and tomography.
\newblock {\em Reports on Progress in Physics}, 73(7):076701, 2010.

\bibitem{american2007american}
American National~Standards Institute.
\newblock {\em American National Standard for safe use of lasers}.
\newblock Laser Institute of America, 2014.

\bibitem{patterson1995absorption}
Michael~S Patterson, Stefan Andersson-Engels, Brian~C Wilson, and Ernest~K
  Osei.
\newblock Absorption spectroscopy in tissue-simulating materials: a theoretical
  and experimental study of photon paths.
\newblock {\em Applied optics}, 34(1):22--30, 1995.

\bibitem{jonsson2020multi}
Joakim J{\"o}nsson and Edouard Berrocal.
\newblock Multi-scattering software: part i: online accelerated monte carlo
  simulation of light transport through scattering media.
\newblock {\em Optics Express}, 28(25):37612--37638, 2020.

\bibitem{Frantz2021}
Joakim D.~Frantz, J{\"o}nsson and Edouard Berrocal.
\newblock Multi-scattering software: part ii: Experimental validation for the
  light intensity distribution.
\newblock {\em Optics Express, article under review}, 2021.

\bibitem{jonsson2021multi}
Joakim J{\"o}nsson.
\newblock {\em Multi-Scattering: Computational light transport in turbid
  media}.
\newblock PhD thesis, Lund University, 2021.

\bibitem{mecklenbrauker1997wigner}
Wolfgang Mecklenbr{\"a}uker and Franz Hlawatsch.
\newblock {\em The Wigner distribution: theory and applications in signal
  processing}.
\newblock Elsevier Science, 1997.

\bibitem{goodman2015statistical}
Joseph~W Goodman.
\newblock {\em Statistical optics}.
\newblock John Wiley \& Sons, 2015.

\bibitem{zhu2018image}
Bo~Zhu, Jeremiah~Z Liu, Stephen~F Cauley, Bruce~R Rosen, and Matthew~S Rosen.
\newblock Image reconstruction by domain-transform manifold learning.
\newblock {\em Nature}, 555(7697):487--492, 2018.

\bibitem{correia2011split}
Teresa Correia, Juan Aguirre, Alejandro Sisniega, Judit Chamorro-Servent, Juan
  Abascal, Juan~J Vaquero, Manuel Desco, Ville Kolehmainen, and Simon Arridge.
\newblock Split operator method for fluorescence diffuse optical tomography
  using anisotropic diffusion regularisation with prior anatomical information.
\newblock {\em Biomedical optics express}, 2(9):2632--2648, 2011.

\bibitem{xu2019improving}
Shiqi Xu, KM~Shihab Uddin, and Quing Zhu.
\newblock Improving dot reconstruction with a born iterative method and
  us-guided sparse regularization.
\newblock {\em Biomedical optics express}, 10(5):2528--2541, 2019.

\bibitem{glorot2010understanding}
Xavier Glorot and Yoshua Bengio.
\newblock Understanding the difficulty of training deep feedforward neural
  networks.
\newblock In {\em Proceedings of the thirteenth international conference on
  artificial intelligence and statistics}, pages 249--256. JMLR Workshop and
  Conference Proceedings, 2010.

\bibitem{kingma2014adam}
Diederik~P Kingma and Jimmy Ba.
\newblock Adam: A method for stochastic optimization.
\newblock {\em arXiv preprint arXiv:1412.6980}, 2014.

\bibitem{yao2014photoacoustic}
Junjie Yao and Lihong~V Wang.
\newblock Photoacoustic brain imaging: from microscopic to macroscopic scales.
\newblock {\em Neurophotonics}, 1(1):011003, 2014.

\bibitem{wang2004image}
Zhou Wang, Alan~C Bovik, Hamid~R Sheikh, and Eero~P Simoncelli.
\newblock Image quality assessment: from error visibility to structural
  similarity.
\newblock {\em IEEE transactions on image processing}, 13(4):600--612, 2004.

\bibitem{wabnitz2014performance}
Heidrun Wabnitz, Dieter~R Taubert, Mikhail Mazurenka, Oliver Steinkellner,
  Alexander Jelzow, Rainer Macdonald, Daniel Milej, Piotr Sawosz, Micha{\l}
  Kacprzak, Adam Liebert, et~al.
\newblock Performance assessment of time-domain optical brain imagers, part 1:
  basic instrumental performance protocol.
\newblock {\em Journal of Biomedical Optics}, 19(8):086010, 2014.

\bibitem{kleiser2018comparison}
S~Kleiser, D~Ostojic, B~Andresen, Nassim Nasseri, H~Isler, Felix Scholkmann,
  T~Karen, G~Greisen, and M~Wolf.
\newblock Comparison of tissue oximeters on a liquid phantom with adjustable
  optical properties: an extension.
\newblock {\em Biomedical optics express}, 9(1):86--101, 2018.

\bibitem{sutin2016time}
Jason Sutin, Bernhard Zimmerman, Danil Tyulmankov, Davide Tamborini, Kuan~Cheng
  Wu, Juliette Selb, Angelo Gulinatti, Ivan Rech, Alberto Tosi, David~A Boas,
  et~al.
\newblock Time-domain diffuse correlation spectroscopy.
\newblock {\em Optica}, 3(9):1006--1013, 2016.

\bibitem{kholiqov2020time}
Oybek Kholiqov, Wenjun Zhou, Tingwei Zhang, VN~Du~Le, and Vivek~J Srinivasan.
\newblock Time-of-flight resolved light field fluctuations reveal deep human
  tissue physiology.
\newblock {\em Nature communications}, 11(1):1--15, 2020.

\bibitem{snell2017prototypical}
Jake Snell, Kevin Swersky, and Richard~S Zemel.
\newblock Prototypical networks for few-shot learning.
\newblock {\em arXiv preprint arXiv:1703.05175}, 2017.

\bibitem{mandt2017stochastic}
Stephan Mandt, Matthew~D Hoffman, and David~M Blei.
\newblock Stochastic gradient descent as approximate bayesian inference.
\newblock {\em arXiv preprint arXiv:1704.04289}, 2017.

\bibitem{lemieux1999investigating}
P-A Lemieux and DJ~Durian.
\newblock Investigating non-gaussian scattering processes by using nth-order
  intensity correlation functions.
\newblock {\em JOSA A}, 16(7):1651--1664, 1999.

\bibitem{jazani2019alternative}
Sina Jazani, Ioannis Sgouralis, Omer~M Shafraz, Marcia Levitus, Sanjeevi
  Sivasankar, and Steve Press{\'e}.
\newblock An alternative framework for fluorescence correlation spectroscopy.
\newblock {\em Nature communications}, 10(1):1--10, 2019.

\end{thebibliography}



\end{document}